\documentclass[12pt,titlepage]{article}
\usepackage{amsmath}
\usepackage[portuguese]{babel}
\usepackage[font=small,format=plain,labelfont=bf,up,textfont=it,up]{caption}
\usepackage{graphicx}

\begin{document}

\title{O exemplo mais simples \\do uso do m\'{e}todo das imagens\footnote {To appear in Revista Brasileira de Ensino de F\'{\i}sica} \\
{\small The simplest example of the use of the method of images}}
\date{}
\author{Antonio S. de Castro\thanks{%
E-mail: castro@pq.cnpq.br} \\
\\
Departamento de F\'{\i}sica e Qu\'{\i}mica, \\
Universidade Estadual Paulista \textquotedblleft J\'{u}lio de Mesquita
Filho\textquotedblright, \\
Guaratinguet\'{a}, SP, Brasil}
\maketitle

\begin{abstract}
Mostra-se que o sistema constitu\'{\i}do por dois planos infinitos e
paralelos, sendo um plano condutor, e o outro um plano isolante com carga
distribu\'{\i}da uniformemente, \'{e} o exemplo mais simples do uso do m\'{e}%
todo das imagens, em constraste com o que \'{e} difundido na literatura.
\newline
\newline
\noindent Palavras-chave: m\'{e}todo das imagens, equa\c{c}\~{a}o de
Poisson, carga induzida.\newline
\newline
\newline
\newline
\newline
{\small \noindent It is shown that the system consisting of two infinite and
parallel planes, one of them a conductor, and the other an insulator with a
uniform distribution of charges, is the simplest example of the use of the
method of images, in contrast to what is diffused in the literature. }
\end{abstract}


\section{Introdu\c{c}\~{a}o}

Em eletrost\'{a}tica, o potencial el\'{e}trico gerado por uma prescrita
distribui\c{c}\~{a}o de cargas na presen\c{c}a de condutores n\~{a}o \'{e}
um problema que possa ser resolvido por meio da integral%
\begin{equation}
V(\overrightarrow{r})=\frac{1}{4\pi \varepsilon _{0}}\int \frac{dq^{\prime }%
}{|\overrightarrow{r}-\overrightarrow{r}^{\prime }|}  \label{p}
\end{equation}%
Isto se d\'{a} porque $dq^{\prime }$ n\~{a}o \'{e} somente devida \`{a} dada
distribui\c{c}\~{a}o de cargas, mas tamb\'{e}m \`{a} distribui\c{c}\~{a}o de
cargas induzida nas superf\'{\i}cies dos condutores, sendo esta desconhecida
\textit{a priori}. Contudo, o problema pode ser resolvido por meio da equa%
\c{c}\~{a}o de Poisson%
\begin{equation}
\nabla ^{2}V(\overrightarrow{r})=-\frac{\rho (\overrightarrow{r})}{%
\varepsilon _{0}}  \label{poi}
\end{equation}%
acrescida de condi\c{c}\~{o}es de contorno. O potencial \'{e} uma fun\c{c}%
\~{a}o cont\'{\i}nua, exceto nos pontos em que $\rho $ exibe singularidades
expressas em termos de fun\c{c}\~{o}es delta de Dirac como \'{e} o caso, por
exemplo, de uma carga pontual $q$ localizada em $\overrightarrow{r_{0}}$:%
\begin{equation}
\rho (\overrightarrow{r})=q\delta (\overrightarrow{r}-\overrightarrow{r_{0}})
\label{del}
\end{equation}%
As condi\c{c}\~{o}es de contorno apropriadas s\~{a}o os potenciais ou as
cargas especificados em cada condutor. Assim sendo, o potencial \'{e}
unicamente determinado em todos os pontos do espa\c{c}o. Ent\~{a}o, qualquer
solu\c{c}\~{a}o da equa\c{c}\~{a}o de Poisson que satisfa\c{c}a as condi\c{c}%
\~{o}es de contorno \'{e} solu\c{c}\~{a}o do problema, n\~{a}o importando o m%
\'{e}todo utilizado, sendo cab\'{\i}vel at\'{e} mesmo o recurso \`{a} intui%
\c{c}\~{a}o e \`{a} analogia.

O m\'{e}todo das imagens \'{e} uma t\'{e}cnica poderosa na resolu\c{c}\~{a}o
de problemas eletrost\'{a}ticos envolvendo uma dada distribui\c{c}\~{a}o de
cargas na presen\c{c}a de condutores. O m\'{e}todo apoia-se na unicidade da
solu\c{c}\~{a}o e consiste na simula\c{c}\~{a}o das condi\c{c}\~{o}es de
contorno pela adi\c{c}\~{a}o de cargas imagens localizadas fora da regi\~{a}%
o de interesse. Contudo, sua efetividade depende da simetria do sistema,
depende da geometria da prescrita distribui\c{c}\~{a}o de cargas e das superf%
\'{\i}cies dos condutores, e assim, como ilustrado nos livros-texto, o m\'{e}%
todo \'{e} restrito a um punhado de sistemas. Landau e Lifshitz \cite{lan}
chegam a mencionar que o mais simples uso do m\'{e}todo das imagens \'{e} a
determina\c{c}\~{a}o do campo el\'{e}trico devido a uma carga pontual nas
proximidades de um meio condutor que ocupa o semiespa\c{c}o. Feynman,
Leighton e Sands \cite{fey} n\~{a}o deixam por menos, e referem-se ao
problema da carga pontual na vizinhan\c{c}a de um plano (folha) condutor
infinito aterrado como sendo a mais simples aplica\c{c}\~{a}o do met\'{o}do
das imagens. \'{E} instrutivo observar que os sistemas mais simples
mencionados nas refer\^{e}ncias \cite{lan} e \cite{fey} s\~{a}o
aparentemente similares, contudo o condutor da Ref. \cite{lan} \'{e}
aterrado por constru\c{c}\~{a}o, enquanto o condutor da Ref. \cite{fey} pode
ter qualquer potencial ou carga prescritos. O problema da carga pontual na
vizinhan\c{c}a de um plano condutor infinito aterrado, como primeiro exemplo
da aplica\c{c}\~{a}o do m\'{e}todo das imagens, \'{e} ub\'{\i}quo na vasta
literatura sobre a eletrodin\^{a}mica cl\'{a}ssica, e Griffths \cite{gri} o
denomina \textquotedblleft o problema cl\'{a}ssico da carga
imagem\textquotedblright .

Neste trabalho, consideramos a aplica\c{c}\~{a}o do uso do m\'{e}todo para
dois planos infinitos e paralelos, sendo um plano condutor, e o outro um
plano isolante com carga distribu\'{\i}da uniformemente, conforme est\'{a}
ilustrado na Figura 1. A bem da verdade, o plano condutor \'{e} uma placa
condutora infinita com superf\'{\i}cies planas (a espessura da placa \'{e}
irrelevante). Mostramos, ent\~{a}o, que os sistemas mais simples mencionados
nas refer\^{e}ncias \cite{lan} e \cite{fey} s\~{a}o t\~{a}o somente os mais
simples dentre todos aqueles outros patentes nos livros-texto. Em nosso
exemplo, tal como no caso da Ref. \cite{fey}, o condutor pode ter qualquer
potencial ou densidade superficial de carga explicitamente antecipados. Come%
\c{c}aremos nossas considera\c{c}\~{o}es com a especifica\c{c}\~{a}o do
potencial do condutor, e depois cuidaremos do condutor carregado.

\section{Condutor com potencial prescrito}

Considere o plano condutor com potencial $V_{0}$ com sua superf\'{\i}cie
superior (superf\'{\i}cie mais pr\'{o}xima ao plano isolante) coincidente
com o plano $XY$ ($z=0$), e o plano isolante carregado uniformemente com
densidade superficial de carga $\sigma $ em $z=d$. Com esta geometria, o
potencial \'{e} independente de $x$ e $y$ de forma que o problema \'{e}
realmente unidimensional: $V=V\left( z\right) $. A regi\~{a}o de interesse
\'{e} a regi\~{a}o $z>0$ e a condi\c{c}\~{a}o de contorno \'{e} $V(0)=V_{0}$%
. A fun\c{c}\~{a}o $V(z)$ \'{e} cont\'{\i}nua apesar da exist\^{e}ncia da
carga superficial em $z=d$. O potencial na regi\~{a}o $z>0$ \'{e} devido a
todas as cargas presentes no sistema. Por\'{e}m, a distribui\c{c}\~{a}o das
cargas induzidas na superf\'{\i}cie superior do condutor n\~{a}o \'{e}
conhecida. Podemos simular o problema real na regi\~{a}o $z>0$ com um plano
isolante carregado uniformemente com densidade superficial de carga $\sigma $
em $z=d$ e um plano isolante carregado uniformemente com densidade
superficial de carga $-\sigma $ em $z=-d$. Neste caso, encontramos%
\begin{equation}
V(z)=\left\{
\begin{array}{cc}
\frac{\sigma }{\varepsilon _{0}}d+V_{0}, & z\geq d \\
&  \\
\frac{\sigma }{\varepsilon _{0}}z+V_{0}, & 0\leq z\leq d%
\end{array}%
\right.   \label{1}
\end{equation}%
Note que $V(0)=V_{0}$ e que o potencial \'{e} cont\'{\i}nuo em $z=d$. O
potencial dado por (\ref{1}) foi obtido pela supress\~{a}o do plano condutor
em $z=0$ e a introdu\c{c}\~{a}o de uma imagem em $z=-d$, mas satisfaz \`{a}
equa\c{c}\~{a}o de Poisson na regi\~{a}o $z>0$ e \`{a} condi\c{c}\~{a}o de
contorno em $z=0$. Ent\~{a}o, devido \`{a} unicidade da solu\c{c}\~{a}o do
problema, o potencial expresso por (\ref{1}) \'{e} a solu\c{c}\~{a}o de
nosso problema original consistindo de um plano condutor com potencial $V_{0}
$ em $z=0$ separado por uma dist\^{a}ncia $d$ de um plano isolante carregado
uniformemente com densidade de carga $\sigma $. Tendo resolvido este
problema, podemos agora calcular todas as grandezas f\'{\i}sicas relevantes,
tais como o campo el\'{e}trico, a carga induzida no condutor, a for\c{c}a
por unidade de \'{a}rea entre as duas distribui\c{c}\~{o}es de carga, a
energia potencial por unidade de \'{a}rea acumulada no sistema, e \textit{%
etecetera}. A prop\'{o}sito, o campo el\'{e}trico $\overrightarrow{E}=\left(
-dV/dz\right) \widehat{k}$ \'{e} expresso por%
\begin{equation}
\overrightarrow{E}(z)=\left\{
\begin{array}{cc}
\overrightarrow{0}, & z>d \\
&  \\
-\frac{\sigma }{\varepsilon _{0}}\,\widehat{k}, & 0<z<d%
\end{array}%
\right.   \label{2}
\end{equation}%
Lembrando que o campo el\'{e}trico em pontos infinitamente pr\'{o}ximos \`{a}
superf\'{\i}cie de um condutor \'{e} $\overrightarrow{E}_{c}=\left( \sigma
_{c}/\varepsilon _{0}\right) \widehat{n}$, onde $\sigma _{c}$ \'{e} a
densidade de carga do condutor e $\widehat{n}$ \'{e} um vetor unit\'{a}rio
perpendicular \`{a} superf\'{\i}cie, podemos concluir que a distribui\c{c}%
\~{a}o de cargas induzidas na superf\'{\i}cie superior do condutor \'{e} $%
-\sigma $.

\section{Condutor com densidade superficial de carga prescrita}

Em vez da especifica\c{c}\~{a}o do potencial do condutor, bem que poder\'{\i}%
amos ter especificado sua densidade de carga $\sigma _{c}$. Neste caso,
podemos aproveitar o resultado da se\c{c}\~{a}o anterior e escrever%
\begin{equation}
V(z)=\left\{
\begin{array}{cc}
-\frac{\sigma _{c}+\sigma }{2\varepsilon _{0}}z+\frac{\sigma }{\varepsilon
_{0}}d+\tilde{V}_{0}, & z\geq d \\
&  \\
-\frac{\sigma _{c}-\sigma }{2\varepsilon _{0}}z+\tilde{V}_{0}, & 0\leq z\leq
d%
\end{array}%
\right.   \label{p2}
\end{equation}%
s\'{o} que desta vez o potencial do condutor ($\tilde{V}_{0}$) \'{e} uma
constante desconhecida. A parcela $\sigma d/\varepsilon _{0}$ presente no
potencial na regi\~{a}o $z\geq d$ assegura a continuidade de $V(z)$ em $z=d$%
. At\'{e} o momento a distribui\c{c}\~{a}o de cargas nas superf\'{\i}cies do
condutor ainda n\~{a}o entrou na hist\'{o}ria. De (\ref{p2}) conclu\'{\i}mos
que o campo el\'{e}trico na regi\~{a}o $z>0$ \'{e} expresso por%
\begin{equation}
\overrightarrow{E}(z)=\left\{
\begin{array}{cc}
\frac{\sigma _{c}+\sigma }{2\varepsilon _{0}}\,\widehat{k}, & z>d \\
&  \\
\frac{\sigma _{c}-\sigma }{2\varepsilon _{0}}\,\widehat{k}, & 0<z<d%
\end{array}%
\right.
\end{equation}%
Da\'{\i}, lembrando mais uma vez a express\~{a}o do campo el\'{e}trico em
pontos infinitamente pr\'{o}ximos \`{a} superf\'{\i}cie de um condutor,
conclu\'{\i}mos que a distribui\c{c}\~{a}o de cargas induzidas na superf%
\'{\i}cie superior do condutor (superf\'{\i}cie mais pr\'{o}xima ao plano
isolante) \'{e} $\left( \sigma _{c}-\sigma \right) /2$. Podemos ent\~{a}o
afirmar que uma carga com densidade $-\sigma $ se distribui na superf\'{\i}%
cie superior do condutor. A carga restante, com densidade superfial $\sigma
_{c}+\sigma $ se distribui igualmente entre as superf\'{\i}cies superior e
inferior do condutor. Em suma, h\'{a} uma carga com densidade $\left( \sigma
_{c}-\sigma \right) /2$ na superf\'{\i}cie superior do condutor e \ uma
carga com densidade $\left( \sigma _{c}+\sigma \right) /2$ em sua superf%
\'{\i}cie inferior.

\section{Coment\'{a}rios finais}

Embora simples em princ\'{\i}pio, o m\'{e}todo das imagens pode se tornar um
tanto complicado para outras geometrias e, neste contexto, destaco o coment%
\'{a}rio de Feynman, Leighton e Sands na p\'{a}gina 6.9 da Ref. \cite{fey}
(tradu\c{c}\~{a}o livre):

\begin{quote}
Nos livros, pode-se encontrar listas longas de solu\c{c}\~{o}es para
condutores de formas hiperb\'{o}licas e outras coisas aparentemente
complicadas, e voc\^{e} se pergunta como algu\'{e}m pode ter resolvido essas
formas terr\'{\i}veis. Elas foram resolvidas ao contr\'{a}rio! Algu\'{e}m
resolveu um problema simples com cargas prescritas. Ent\~{a}o viu que alguma
superf\'{\i}cie equipotencial apareceu em uma nova forma, e escreveu um
trabalho no qual salientou que o campo fora dessa forma particular pode ser
descrito de uma certa maneira.
\end{quote}

Esse coment\'{a}rio sarc\'{a}stico poderia ser imputado \`{a} configura\c{c}%
\~{a}o adotada neste trabalho, haja vista que poder\'{\i}amos ter obtido
todas as nossas conclus\~{o}es usando a express\~{a}o para o potencial
gerado por um plano isolante infinito com carga distribu\'{\i}da
uniformemente, e apelando simplesmente \`{a} nulidade do campo el\'{e}trico
no interior do condutor. \'{E} ineg\'{a}vel, contudo, que os dois casos
abordados neste trabalho revelam-se os mais simples exemplos que podem
ilustrar a aplica\c{c}\~{a}o do m\'{e}todo das imagens com um m\'{\i}nimo de
confus\~{a}o anal\'{\i}tica. O potencial na regi\~{a}o de interesse \'{e} fun%
\c{c}\~{a}o de uma vari\'{a}vel, o campo el\'{e}trico \'{e} secionalmente
uniforme, e a carga induzida no condutor n\~{a}o requer integra\c{c}\~{a}o.

\bigskip

\bigskip

\bigskip

\bigskip

\noindent{\textbf{Agradecimentos}}

O autor \'{e} grato ao CNPq pelo apoio financeiro. Um \'{a}rbitro atencioso
contribuiu para proscrever incorre\c{c}\~{o}es constantes na primeira vers%
\~{a}o deste trabalho.

\begin{figure}[]
\includegraphics{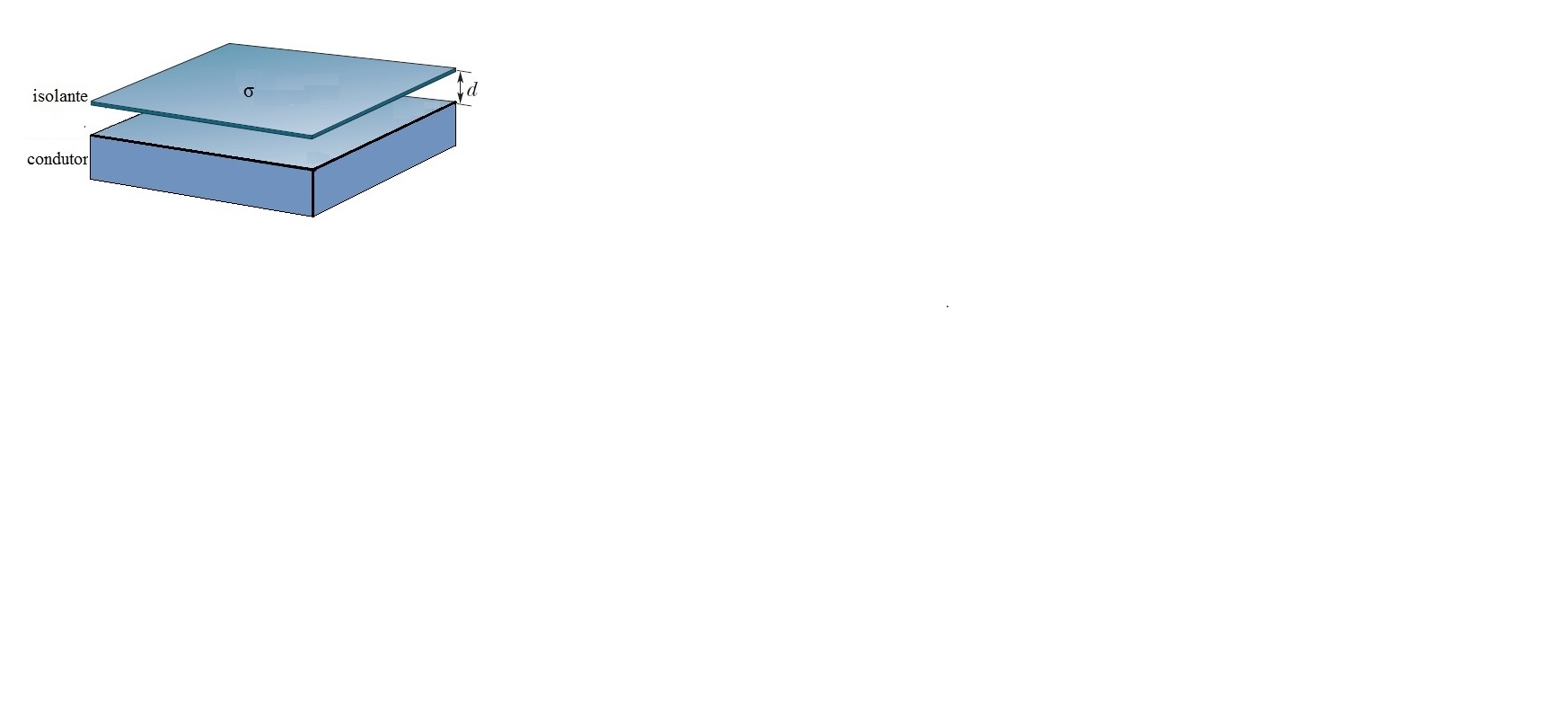}
\caption{Plano isolante com carga distribu\'{\i}da uniformemente ($\protect%
\sigma $) separado pela dist\^{a}ncia $d$ de uma placa condutora infinita
com superf\'{\i}cies planas.}
\label{Fig1}
\end{figure}

\end{document}